# Effect of laser irradiation on the tribological properties of RF-sputtered nickel oxide (NiO) thin films

Srikanth Itapu, Vamsi Borra, Frank X. Li, Pedro Cortes and Mohit Hemanth Kumar

*Abstract—The present work aims at investigating the effect of laser irradiation on the tribological properties of RF-sputtered NiO thin films deposited on industrial grade aluminum substrate. A semiconductor laser based on Nd:YAG operating at its 4th harmonic wavelength, λ = 266nm with varying laser fluence and spot size of about 5 μm is irradiated on the NiO film. The localized heating allows for smoothening of the NiO film along with contributions to the changes in the stoichiometry of NiO (reduction of excess oxygen). In particular, the effects of tuning laser fluence and the subsequent tribology tests pertaining to the coefficient of friction variations for tribological tests are discussed.*

## I. INTRODUCTION

Surface coatings serve as great substitutes to liquid lubricants in enhancing the properties of materials in extremely harsh environments. Surface coatings are indeed prescribed for tribological applications which are less toxic as well as have stability with respect to chemical compositions. Such coating can function as lubricants and mitigate surface damage. Thin films improve coefficient of friction (COF) as well as wear resistance [1-2]. The transition metal oxides play a pivotal part because of their high stability at elevated temperatures and in reactive environments. It is also widely understood for literature that metal nanoparticles such as Nickel (Ni) and Copper (Cu) nanoparticles form low shear strength films on the surface of friction pairs. Compared with traditional organic long-chain additives, the use of metal nanoparticles has simple compositions in the friction process and the tribological properties of lubricants were greatly improved [3-5].

In this regard, Zinc Oxide (ZnO) thin films have found widespread tribological because of its low thermal expansion, and high melting temperature. Due to negligible variations in hardness values in ZnO films fabricated at various RF power levels, it was proven difficult to make an exact correlation between the tribology properties and Vickers hardness of films [6]. Greater lubricating property was proposed by authors using adaptive-lubrication of Molybdenum in the temperature range of 20–700 °C. The amount of Molybdenum Oxide ($MoO_3$) in the reaction was determined by the content of carbon in the molten pool. The wear experiments conducted up to 800 °C evaluated enhanced tribological properties of $MoO_3$ [7].

The tribological properties of $Al_2O_3$ coatings electrodeposited from the Watts bath contained various amounts of $Al_2O_3$ nanoparticles. Thus, the friction and wear resistance, were investigated not in entirety yet. The improved microhardness due to the increased $Al_2O_3$ content in the deposited layer was highly correlated with the volume fraction of alumina [8].

The composite coatings with nanostructured $TiO_2$ also exhibit excellent tribological behavior. Researchers in [9] had investigated the influence of heat treatment on tribological behavior and mechanical characteristics. Their results revealed agglomerated spheroids and porous morphology, which act as a pivot in the enhancing $TiO_2$.

Higher temperatures essentially impact the mechanical and friction properties of the components, even causing fatigue cracks. Laser cladding was proposed as an efficient method for composite fabrication forming metallurgical bonding to the substrate [10]. Significant upfront cost to setup the laser cladding systems renders the method ineffective. In this regard, the proposed work on using pulsed-laser irradiation technique for enhancing the tribological properties of transition metal oxide thin films offer the advantage of localized thermal heating, noncontact nature as compared to laser cladding [11-13].

Demonstrating industrial compatibility of transition metal oxide depositions is essential and include sputtering, electro-deposition and vacuum evaporation. Sputtering has the advantage of depositing a wide range of thin films materials and NiO is one particularly interesting material due to its wide range of applicability [12], [14]. The

*Corresponding authors:

Srikanth Itapu (srikanth.itapu@alliance.edu.in) and Vamsi Borra (vsborra@ysu.edu)

F. X. Li, V. Borra are with Electrical and Computer Engineering Program, College of STEM, Rayen School of Engineering, Youngstown State University, Youngstown, Ohio, 44555 USA (corresponding author e-mail: vsborra@ysu.edu).

S. Itapu is with Department of Electronics & Communication Engineering, Alliance University, Bengaluru, 562106 India (corresponding author e-mail: srikanth.itapu@alliance.edu.in).

P. Cortes is with Chemical Engineering Program, College of STEM, Rayen School of Engineering, Youngstown State University, Youngstown, Ohio, 44555 USA.

M.H. Kumar is with Department of Mechanical Engineering, Alliance University, Bengaluru, 562106 India.

stoichiometric NiO is a Mott insulator with a conductivity of $10^{-13}$ S/cm, while nonstoichiometric NiO is a wide-band-gap p-type semiconductor. Although structural and electrical properties of NiO films have been studied extensively, mechanical and tribological properties have been recently trending primarily due its ease of deposition. Therefore, understanding the mechanical correlations with micropatterning of NiO-based films has been of great interest. It is conceived that the type of deposition of thin films has a direct effect on the film micropatterning and stoichiometry. Ni-based metal matrix composites (MMCs) are widely used as low-cost solid lubricants with good wear performance under high temperature conditions in addition to the existing electronic, optical, magnetic applications. For example, Ni-based composites are employed to enhance surface quality, such as cutting tools, turbine engine components, and wearing plates.

The present work aims at investigating post-deposition laser irradiation effects on surface wettability of NiO thin films deposited on Aluminum substrates by radio-frequency magnetron sputtering. A semiconductor laser based on Nd:YAG operating at its 4th harmonic wavelength, $\lambda = 266$ nm with varying laser fluence and spot size of about 5 µm is irradiated on the NiO film. The localized heating allows for smoothening of the NiO film along with contributions to the changes in the stoichiometry of NiO (reduction of excess oxygen) and creation of Ni interstitials. In particular, the effects of tuning laser fluence and the subsequent film microstructuring and the associated nanomechanical properties of the NiO thin films revealed by the mechanical tests, wear and coefficient of friction variations for tribological tests are discussed.

## II. SAMPLE PREPARATION

NiO thin films were fabricated on aluminum substrate by an RF magnetron sputtering. In this work, the Ni target (99.9% Ni) was used as the target in the presence of argon/oxygen (Ar/O$_2$) with a gas flow of 4:1 to realize a stoichiometric NiO film. The optimized NiO coating of thickness 100nm was measured using a Dektak Profilometer. The parameters for deposition of NiO thin films are mentioned in table 1. The aluminum plate surface was mechanically polished using an automatic grinding machine. The ground sample was again polished using a liquid suspension containing 0.25 µm size silicon carbide nanoparticles, to get a smoother appearance. The polished plates were then cleaned for 5 mins in an ultrasonic cleaner, rinsed with distilled water, sprayed with ethanol, and dried under a heat gun.

Laser irradiation setup is discussed elsewhere and shown in figure 1 [15, 16]. The COF was measured using a pin-on-disc tribometer. The thin film deposited substrate was mounted on the chuck and the pin was placed on the surface of the disk. A detailed description of this instrument can be found elsewhere [17, 18]. The substrate has dimensions of 6 cm x 6 cm, thickness 0.8 mm and polished to remove any surface cracks. It was then cleaned with De-ionized (DI) water, and isopropyl alcohol to remove any unwanted dust. Finally, the substrates were then ultrasonicated in distilled water. The pin-on-disc tribometer tests were conducted on laser irradiated and as-deposited samples. The contact angle measurements for laser irradiated NiO film with various laser fluences were performed using 1µl distilled water and having a stabilization time of 10 seconds.

TABLE I. PARAMETERS FOR DEPOSITION OF NiO THIN FILMS.

| Target | 3-in diameter nickel metal (99.99% pure) |
|---|---|
| Substrate | Aluminum |
| Sputtering pressure | 8 mTorr |
| Sputtering gas | (Ar:O$_2$) - 20:80 |
| RF power | 125 W |
| Target to substrate distance | 8 cm |
| Laser and operating wavelength | Nd:YAG (pulsed), $\lambda = 266$ nm |
| Film thickness | 100 nm |

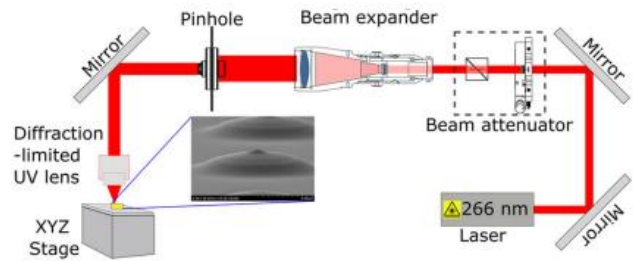

**Figure 1.** Experimental setup for laser irradiation on NiO films [15, 16].

## III. RESULTS AND DISCUSSION

Figure 2 represents the Scanning Electron Microscopy (SEM) images of the Laser-irradiated NiO on Ni substrate with varying laser fluences. The localized heating due to laser irradiation of 0.35 J/cm$^2$ created a crater on the NiO film with debris flying around the vicinity of the laser spot. As the laser fluence is increased to 0.4 J/cm$^2$, the molten NiO holds itself and smoothening of the film is observed. A further increase in the laser fluence to 0.45 J/cm$^2$ results in deeper crater formations, suggesting that a part of the laser energy is transferred to the Ni substrate.

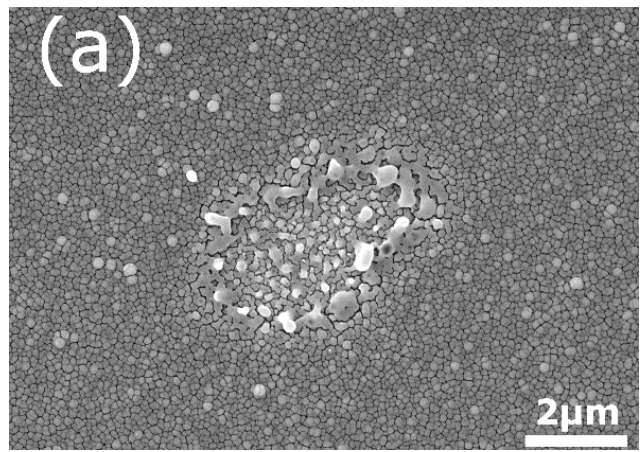

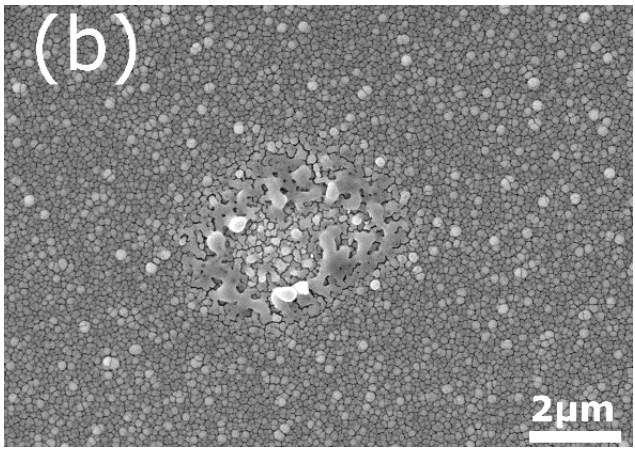

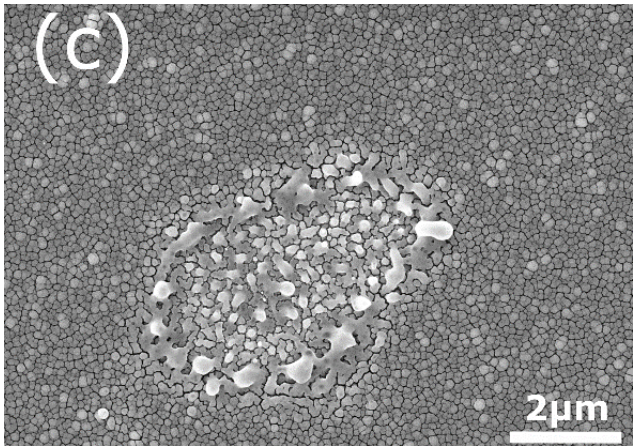

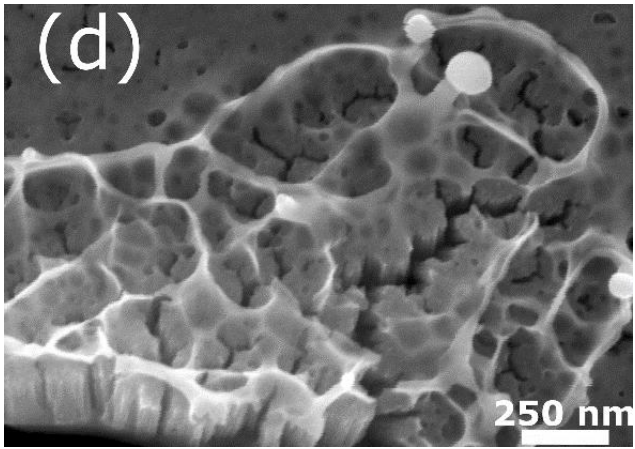

**Figure 2.** SEM images of Laser irradiation on NiO thin film deposited on Ni substrate (a) laser fluence of 0.35 J/cm$^2$, (b) 0.4 J/cm$^2$, (c) 0.45 J/cm$^2$ and (d) magnified SEM image of molten area due to laser irradiation on 0.4 J/cm$^2$.

Table 2 summarizes the Vickers hardness of the laser irradiated NiO film and the aluminum substrate. Noticeable difference in hardness was observed because laser irradiation results in localized melting of NiO and thus decreases the hardness of the NiO coating/aluminum substrate. This decrease in hardness increased the ductility of the NiO film which cause it to wear faster.

TABLE II. VICKERS HARDNESS OF THE SUBSTRATE AND LASER-IRRADIATED NiO

| Material | Vicker's harness (MPa) |
|---|---|
| Aluminum substrate | 38.6 |
| NiO (0.35 J/cm$^2$) | 18.7 |
| NiO (0.40 J/cm$^2$) | 18.5 |
| NiO (0.45 J/cm$^2$) | 18.2 |

Energy dispersive X-ray spectroscopy (EDS) is performed using a Hitachi field-emission scanning electron microscope (FESEM) F-4800 to check the chemical composition of the NiO films before and after laser irradiation. Table 3 represents the composition of as-deposited and laser irradiated NiO films.

TABLE III. COMPOSIITON OF Ni AND O FOR LASER-IRRADIATED NiO

| Sample type | Ni % | O % | O:Ni ratio |
|---|---|---|---|
| As-deposited | 41.06 | 59.94 | 1.45 |
| NiO (0.35 J/cm$^2$) | 44.98 | 55.02 | 1.22 |
| NiO (0.40 J/cm$^2$) | 48.04 | 51.96 | 1.08 |
| NiO (0.45 J/cm$^2$) | 51.41 | 48.59 | 0.94 |

The COF for various laser fluences of the NiO coatings was compared. The friction coefficient ($\mu_k$) was calculated for steady state kinetic friction. A constant speed of 100 rpm and constant normal load of 0.25 N were set to perform the friction measurements. The experiments were conducted at ambient room temperature and pressure.

Figure 3 represents the effect of laser irradiation on the co-efficient of friction of NiO films deposited on Al substrate. With reference to figure 1(d), the molten NiO for 0.4 J/cm$^2$ laser fluence causes the friction to reduce significantly dropping from 0.6 for Al substrate alone to 0.22 for NiO/Al substrate. A slight deviaition from the optimised COF for 0.45 J/cm$^2$ laser fluence is observed. This may be attributed to laser energy penetrating into the substrate , thus creating roughness in the Al substrate.

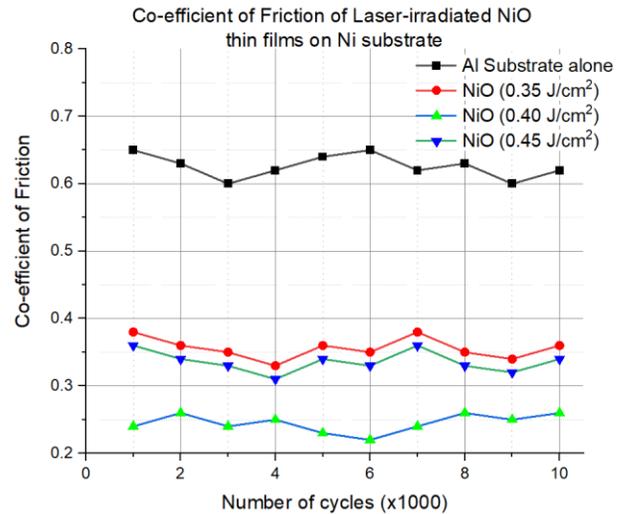

**Figure 3**. The co-efficient of fricition of laser-irradiated NiO film on Ni substrate at 100 rpm.

Figure 4 represents the contact angle measurements for the as-deposited and laser irradiated NiO. The left and right contact angles are obtained as $95.7^0$ for a laser fluence of 0.45 J/cm$^2$, which suggests that the laser irradiated NiO film on Al substrate is hydrophobic in nature.

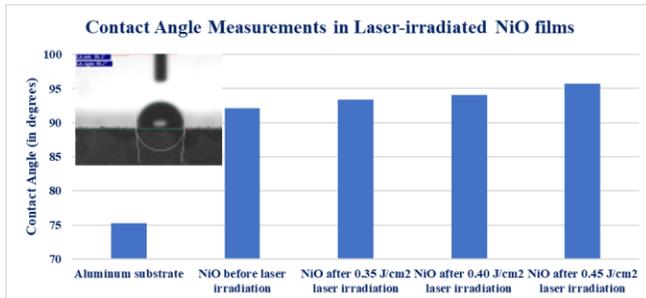

**Figure 4.** Contact angle measurements for laser irradiated NiO thin films.

## IV. CONCLUSION

This work establishes a dependence of laser fluence on the tribolgoical properties of NiO films depsoited by RF sputtering on aluminum substrate. The localized melting due to optimized laser fluence resulted in molten NiO (as a result, reduction in any excess oxygen), thus smoothening the NiO/Al. It also renders to be have low co-effiecient of friction as compared to the Al substrate alone.